\documentclass[letterpaper, 10 pt, conference]{ieeeconf}  % Comment this line out if you need a4paper

\IEEEoverridecommandlockouts                              

\usepackage{graphics} % for pdf, bitmapped graphics files
\usepackage{graphicx}
\usepackage{epsfig} % for postscript graphics files
\usepackage{mathptmx} % assumes new font selection scheme installed
\usepackage{times} % assumes new font selection scheme installed
\usepackage{amsmath} % assumes amsmath package installed
\usepackage{amssymb}  % assumes amsmath package installed

\title{\LARGE \bf
Wearable-Based Real-time Freezing of Gait Detection in Parkinson’s Disease Using Self-Supervised Learning}

\author{Shovito Barua Soumma$^{1}$, Kartik Mangipudi$^{2}$, Daniel Peterson$^{1}$, Shyamal Mehta$^{3}$ and Hassan Ghasemzadeh$^{1}$
\thanks{$^{1}$Shovito Barua Soumma, Daniel S. Peterson, and Hassan
Ghasemzadeh are with the College of Health
Solutions, Arizona State University, Phoenix, AZ 85004, USA. Email: \{shovito, daniel.peterson1,
hassan.ghasemzadeh\}@asu.edu.}
\thanks{$^{2}$Kartik Mangipudi is an Assistant Professor of Neurology at the University of Florida, USA. Email:{kartik.mangipudi@jax.ufl.edu}. $^{3}$Shyamal Mehta is an Associate Professor of Neurology at the Mayo Clinic, Arizona, USA. Email: {mehta.shyamal@mayo.edu}.}%
}

\begin{document}

\maketitle
\thispagestyle{empty}
\pagestyle{empty}

\begin{abstract}
LIFT-PD is an innovative self-supervised learning framework developed for real-time detection of Freezing of Gait (FoG) in Parkinson’s Disease (PD) patients, using a single triaxial accelerometer. It minimizes the reliance on large labeled datasets by applying a Differential Hopping Windowing Technique (DHWT) to address imbalanced data during training. Additionally, an Opportunistic Inference Module is used to reduce energy consumption by activating the model only during active movement periods. Extensive testing on publicly available datasets showed that LIFT-PD improved precision by 7.25\% and accuracy by 4.4\% compared to supervised models, while using 40\% fewer labeled samples and reducing inference time by 67\%. These findings make LIFT-PD a highly practical and energy-efficient solution for continuous, in-home monitoring of PD patients.
\end{abstract}

\section{Introduction}
Parkinson’s Disease (PD) is a progressive neurodegenerative disorder that significantly affects motor functions. Freezing of Gait (FoG) is one of the most disabling motor symptoms, where patients experience sudden and brief periods of immobility, leading to a heightened risk of falls. Early detection and continuous FoG monitoring are critical for timely interventions and improving the quality of life for PD patients. However, existing detection systems often require extensive labeled data, multiple sensors and high computational resources, making them less suitable for real-world, continuous monitoring using wearable devices~\cite{sigcha2022improvement, naghavi2021towards, mikos2017real, soumma2024mds}. To address these challenges, we present an innovative label-efficient, patient-independent, and robust self-supervised learning framework, \textit{\textbf{LIFT-PD} (\textbf{L}abel-efficient \textbf{I}n-home \textbf{F}reezing-of-Gait \textbf{T}racking in \textbf{P}arkinson’s \textbf{D}isease)}, for detecting FoG events in real time. 
\section{Proposed Method}

\begin{figure}[h]
\centering
\includegraphics[width=1\linewidth,trim={1 8 30 2},clip]{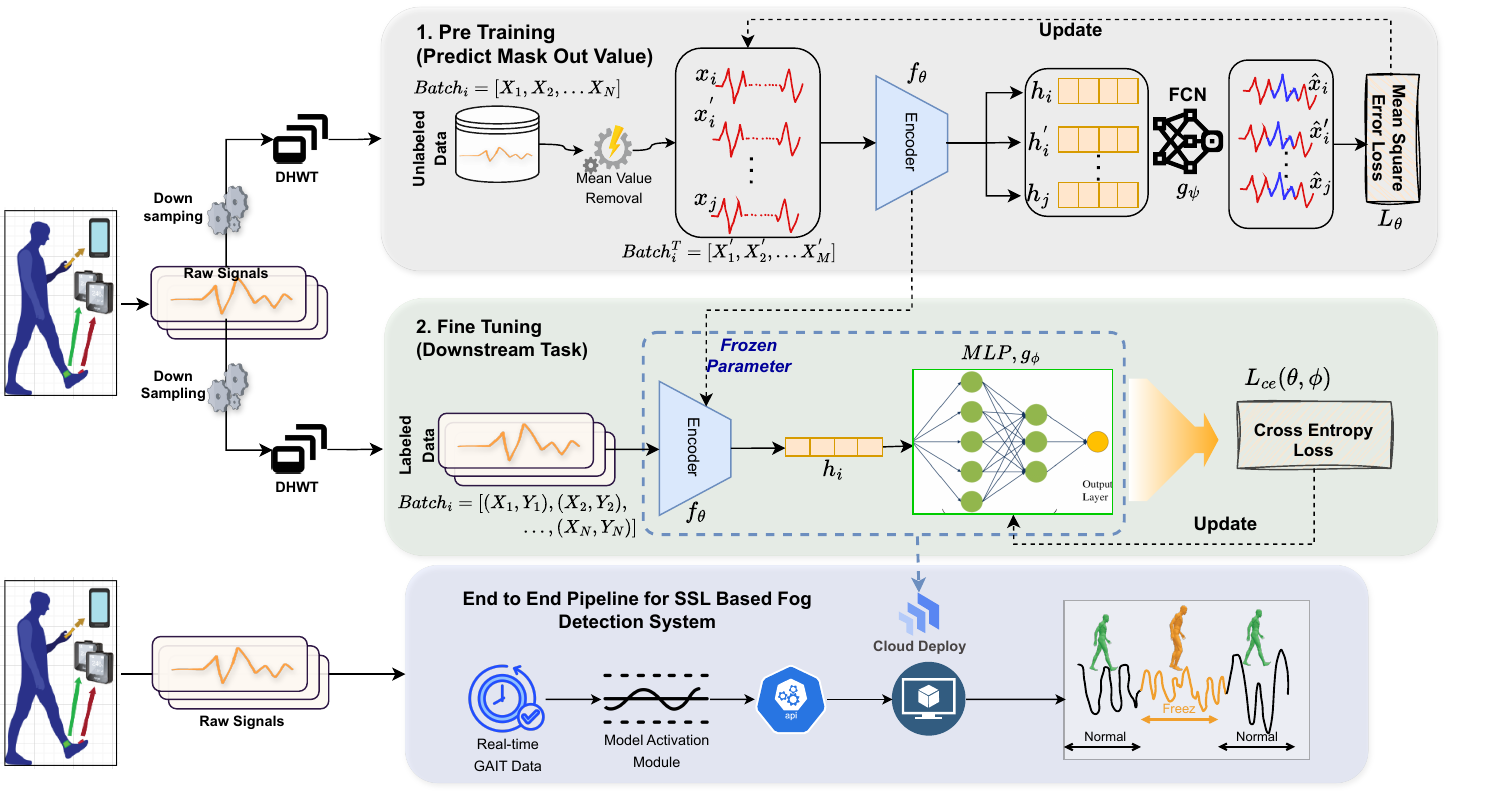}

\caption{{Self-supervised training pipeline.
\textbf{Pretrain:} Encoder reconstructs masked signal segments, outputting predicted values \(h_i\) for missing data.
\textbf{Fine-tuning:} Encoder weights are frozen while the MLP is optimized using cross-entropy loss on labeled data. \textbf{\textit{Model Activation Module}}  selectively activates the computationally intensive FoG detection model.}}
\label{fig:method}
\end{figure}

LIFT-PD is a novel self-supervised learning (SSL) framework designed to detect FoG episodes in real-time without the need for large labeled datasets. LIFT-PD incorporates an unique Differential Hopping Windowing Technique (DHWT) to balance imbalanced datasets during training, thus improving the model’s ability to generalize across diverse patient populations. To enhance power efficiency, LIFT-PD includes an Opportunistic Inference Module that selectively activates the model only during active periods of patient movement, significantly reducing energy consumption. The model was validated on multiple publicly available datasets and tested under various patient conditions, including different medication states (On/Off), severities (Mild/Severe), and demographic groups (age, gender).

The FoG event detection problem is framed as a multivariate time-series classification task. At each time stamp \( t \), the input raw signal is represented as a vector
{
\small
{\(\mathbf{x}_t = [x_{t}^{1}, x_{t}^2, x_{t}^3]\)}
}
where 
{\small
\(\mathbf{x}_t \in \mathbb{R}^{c=3} \)
} 
and 
$c$ corresponds to the three-channel (x, y, z). 
These raw signals are then combined into a matrix, 
{\small
$\mathbf{X} = [x_1, x_2, \ldots, x_T] \in \mathbb{R}^{T \times C}$.
} 
After applying the DHWT method, the signals are transformed into \(N\) number of training frames 
{\small
$(\mathbf{X}\rightarrow \mathbf{X_W} \in \mathbb{R}^{N \times T^\prime \times C})$ 
}
where 
{\small
\(x_{w_i} \in X_W\)
} represents $i'$th window and 
{\small
$\mathbf{X_W} = [x_{w_1}, x_{w_2}, \ldots, x_{w_N}]$}. 
The ultimate goal is to correctly assign a label \( y \in \{0,1\} \) to each window.

LIFT-PD uses SSL for FoG detection in two steps: learning contextual representations from raw signals using a 1D CNN model, followed by performing the downstream FoG detection task. In this paper, raw accelerometer signals are used as physiological signals and the downstream task is binary `Freezing of Gait' detection.

\section{Results}
\begin{figure}[h]
\centering
\includegraphics[width=1\linewidth]{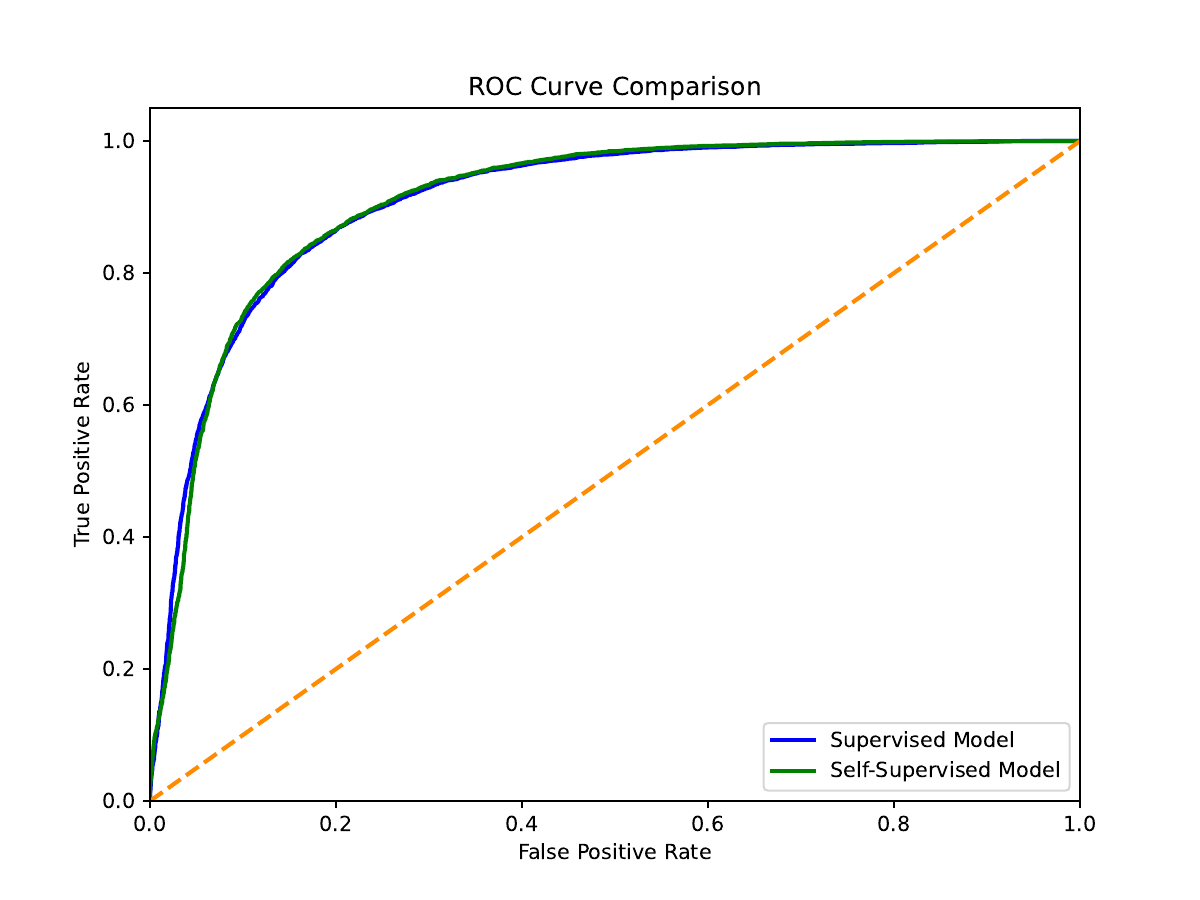}

\caption{Receiver Operating Characteristic (ROC) curves for supervised and self-supervised models}

\label{fig:roc}
\end{figure}

In a previous study ~\cite{10054906}, we employed performance metrics such as sensitivity, specificity, F1 score, and the AUC of receiver operating characteristics (ROC) to evaluate a binary classification model. As this is also a binary classification problem (FoG or non-FoG),
we will use the same metrices here to evaluate our LIFT-PD framework. We extensively
evaluated our proposed system using these metrics, compa-
rable to other state-of-the-art methods.

The ROC curve, shown in Fig.~\ref{fig:roc}, provides a comprehensive evaluation of the classification performance for both models.  A larger area under the ROC curve indicates better classification performance. The supervised model achieved an AUC of 0.9078, while the SSL model attained a slightly higher AUC of 0.908. The high AUC score of SSL model demonstrates its ability to distinguish between FoG and non-FoG events accurately, making it suitable for real-time FoG detection applications in PD.

Our results demonstrate that LIFT-PD achieves competitive performance compared to state-of-the-art supervised models while \textit{using 40\% fewer labeled samples}. The system delivered 7.25\% higher precision and 4.4\% greater accuracy in detecting FoG episodes, along with a 67\% reduction in inference time due to opportunistic model activation. LIFT-PD consistently outperformed other models in handling imbalanced data and minority classes, such as severe PD cases. 

\section{Conclusions}
Overall, LIFT-PD offers a scalable, energy-efficient, and patient-independent solution for real-time FoG detection using only a single triaxial accelerometer, with reduced power consumption and minimal reliance on labeled data, making it a practical choice for continuous monitoring on wearable devices.
\addtolength{\textheight}{-12cm}   

\bibliographystyle{ieeetr}
\bibliography{sample.bib}

\end{document}